\documentclass[MSNbibl,number,citesort,seceqn,dvips]{arxstspdf}
\usepackage{flushend}
\usepackage{stfloats}
\usepackage{graphicx}

\volume{29}
\issue{1}
\pubyear{2014}
\firstpage{69}
\lastpage{80}
\doi{10.1214/13-STS420} 

\makeatletter
\def\printead@fmt#1{#1}

\makeatother

\begin{document}
\begin{frontmatter}

\title{Search for the Wreckage of Air France Flight AF 447\thanksref{T1}}%
\relateddois{T1}{Discussed in \relateddoi{d}{10.1214/13-STS447} and \relateddoi{d}{10.1214/13-STS463}.}
\runtitle{Search for Air France Flight AF 447}

\begin{aug}
\author[a]{\fnms{Lawrence D.} \snm{Stone}\corref{}\ead[label=e1]{stone@metsci.com}},
\author[a]{\fnms{Colleen M.} \snm{Keller}\ead[label=e2]{keller@metsci.com}},
\author[a]{\fnms{Thomas M.} \snm{Kratzke}\ead[label=e3]{kratzke@metsci.com}\ead[label=u1,url]{www.metsci.com}}
\and
\author[b]{\fnms{Johan P.} \snm{Strumpfer}\ead[label=e4]{Johan.Strumpfer@gmail.com}}
\runauthor{Stone, Keller, Kratzke and Strumpfer}

\affiliation{Metron, Inc., Metron, Inc., Metron, Inc. and University of Cape Town}

\address[a]{Lawrence Stone is Chief Scientist, Colleen Keller and
Thomas Kratzke
are Senior Analysts, Metron, Inc., 1818 Library Street, Suite 600,
Reston, Virginia 20190, USA (e-mail: \printead*{e1};
\printead*{e2}; \printead*{e3}; \printead{u1}).}
\address[b]{Johan Strumpfer is Visiting Professor, Graduate School
of Business, University of Cape Town, Cape Town, South Africa
(\printead{e4}).}

\end{aug}

%
\begin{abstract}
In the early morning hours of June 1, 2009, during a flight from Rio de
Janeiro to Paris, Air France Flight AF 447 disappeared during stormy
weather over a remote part of the Atlantic carrying 228 passengers and
crew to their deaths. After two years of unsuccessful search, the
authors were asked by the French Bureau d'Enqu\^etes et d'Analyses pour
la s\'ecurit\'e de l'aviation to develop a probability distribution for
the location of the wreckage that accounted for all information about
the crash location as well as for previous search efforts.

We used a Bayesian procedure developed for search planning to produce
the posterior target location distribution. This distribution was used
to guide the search in the third year, and the wreckage was found with
one week of undersea search. In this paper we discuss why Bayesian
analysis is ideally suited to solving this problem, review previous
non-Bayesian efforts, and describe the methodology used to produce the
posterior probability distribution for the location of the wreck.
\end{abstract}

%

\end{frontmatter}

\section{Background}
\begin{figure*}

\includegraphics{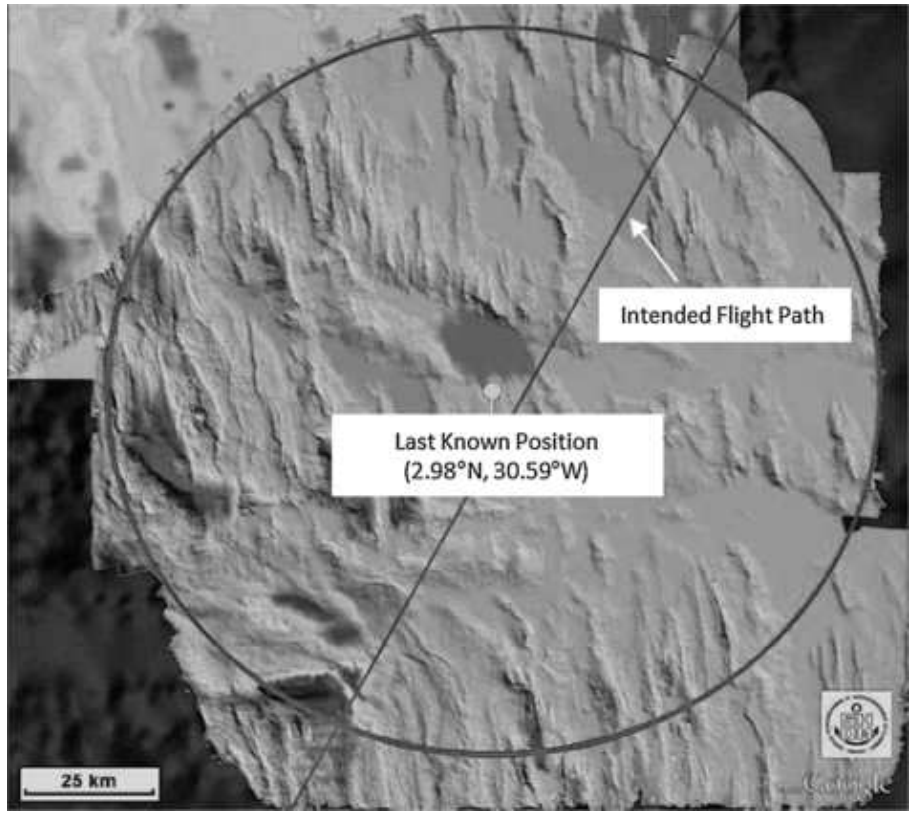}

\caption{Last known position of the aircraft, intended flight path and
the 40 NM circle.}%
\label{figLastKnownPos}
\end{figure*}

In the early morning hours of June 1, 2009, Air France Flight AF 447,
with 228 passengers and crew aboard, disappeared during stormy weather
over the Atlantic while on a flight from Rio de Janeiro to Paris. Upon
receiving notification of the crash, the French Bureau d'Enqu\^etes et
d'Analyses (BEA) pour la s\'ecurit\'e de l'aviation and French search
and rescue authorities organized an international search by aircraft
and surface ships to look for signs of the plane and possible
survivors. On the sixth day of this effort, the first debris and bodies
were found 38 NM north of the aircraft's last known position. That day
a large portion of the galley was found along with other debris and
some bodies. Figure \ref{figLastKnownPos} shows the aircraft's last
known position, intended flight path and a 40 NM circle about the last
known position. Analysis by the BEA determined that the wreckage had to
lie within 40 NM of the plane's last known position.

The aircraft was equipped with a flight data recorder and a cockpit
voice recorder. Each of these recorders was fitted with an underwater
locator beacon that activates an acoustic signal upon contact with
water. The BEA initiated a search to detect these beacons. The search
was performed by two ships employing passive acoustic sensors supplied
by the U.S. Navy and operated by personnel from Phoenix International.
The search began on June 10, 2009 and lasted 31 days until the time
when the batteries in the beacons were estimated to be exhausted. The
ships searched extensively along the intended flight path, but the
beacons were not detected. Next the BEA began an active acoustic search
with side-looking sonar to detect the wreckage on the ocean bottom.
This search took place in August 2009 south of the last known position
in an area not covered by the passive acoustic search. This search was
also unsuccessful.

\begin{figure*}

\includegraphics{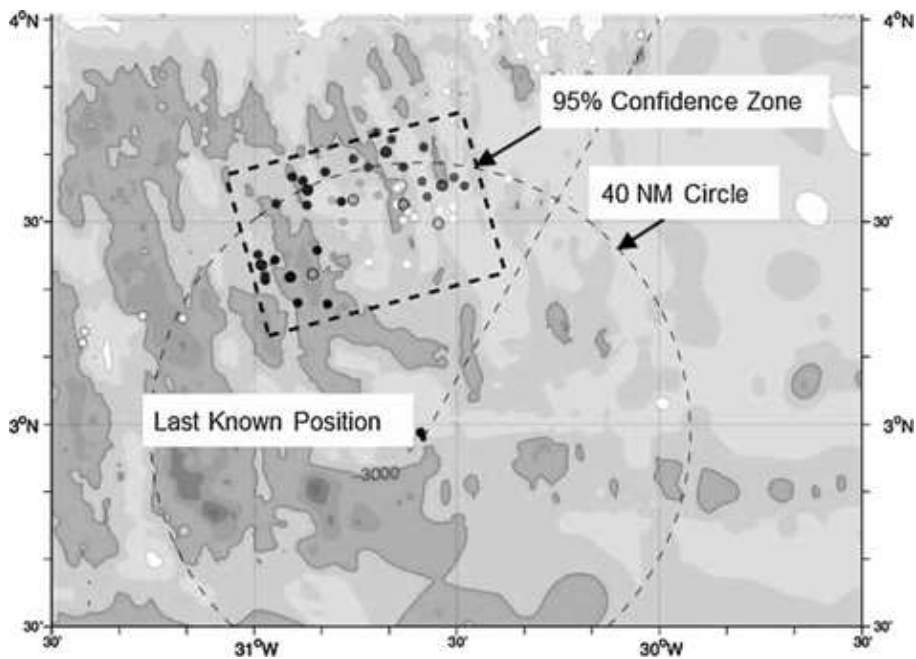}

\caption{The $95\%$ confidence zone recommended in \cite
{Ollitrault2010} for the 2010 search area.}%
\label{figOllitraultSearchArea}%
\end{figure*}

After the unsuccessful search in 2009, the BEA commissioned a group of
oceanographic experts to estimate the currents in the area at the time
of the crash and to use these estimates along with the times and
locations where the surface search had found bodies and debris in order
to estimate the location of the wreckage. In \cite{Ollitrault2010} the
group recommended the rectangular search area north and west of the
last known position shown in Figure \ref{figOllitraultSearchArea}.
This rectangle is described as a $95\%$ confidence zone. The group used
available current measurements to make a number of estimates of the
currents in the area of the wreck at the time of the loss. Using these
estimates, they performed a backward drift on recovered debris and
bodies to produce a number of trajectories ending at an estimated
location of the crash. The group removed trajectories that they felt
were outliers. A bivariate normal error was estimated for each of the
remaining crash location estimates and used to produce a weighted mean
with a bivariate normal error distribution. This error distribution was
used to compute a rectangle centered at the mean with a $95\%$
probability of ``containing'' the wreck location. This rectangle guided
the active acoustic search in April and May of 2010.

The 2010 searches were performed by two teams. The U.S. Navy and
Phoenix International team used a towed side-scan sonar system. The
Woods Hole Oceanographic Institute team used autonomous underwater
vehicles with side-scan sonars and a remotely operated vehicle. After
an unsuccessful search in the rectangle, the teams extended their
efforts to the south and west of the rectangle. Unfortunately, this
search was also unsuccessful.

In July 2010 we were tasked by the BEA to review all information about
the loss of AF 447 as well as the previous search efforts to produce a
probability distribution (map) for the location of the underwater
wreckage. The probability maps that resulted from this process were
used to guide the 2011 search.

On April 3, 2011, almost two years after the loss of the aircraft, the
underwater wreckage was located on the ocean bottom some 14,000 feet
below the surface. On April 8, 2011, the director of the BEA stated,
``This study \cite{StoneEtAl2011BEA} published on the BEA website
on 20 January 2011, indicated a strong possibility for discovery of the
wreckage near the center of the Circle. It was in this area that it was
in fact discovered after one week of exploration'' \cite{Troadec2011}.
Subsequently, the flight data recorder and cockpit voice recorder were
found, retrieved from the ocean bottom and flown to the BEA in Paris
where the data in these recorders were recovered and analyzed. This
data provided crucial information for determining the cause of the
crash. Finding the wreckage also allowed the BEA to return the bodies
of many passengers and crew to their loved ones.

In the sections below we describe the Bayesian process used to compute
these probability distributions.

\section{Why Bayesian Analysis}
Bayesian analysis is ideally suited to planning complicated and
difficult searches involving uncertainties that are quantified by a
combination of objective and subjective probabilities. This approach
has been applied to a number of important and successful searches in
the past, in particular, the searches for the \textit{USS Scorpion}
\cite{RichardsonAndStone1971} and \textit{SS Central America} \cite
{Stone1992}. This approach is the basis for the U.S. Coast
Guard's\vadjust{\goodbreak}
Search and Rescue Optimal Planning System (SAROPS) that is used to plan
Coast Guard maritime searches for people and ships missing at sea \cite
{Fusion2010}.

Complicated searches such as the one for AF 447 are onetime events. We
are not able to recreate the conditions of the crash 1000 times and
record the distribution of locations where the aircraft hits the water.
As a result, definitions of probability distributions in terms of
relative frequencies of events do not apply. Instead, we are faced with
computing a probability distribution on the location of the wreckage
(search object) in the presence of uncertainties and conflicting
information that require the use of subjective probabilities. The
probability distribution on which the search is based is therefore a
subjective one. It is based on the analysts' best understanding of the
uncertainties in the information about the location of the search object.

In an ideal situation, search effort is applied in an optimal fashion
to maximize the probability of detecting the search object within the
effort available. The optimal search problem is a Bayesian decision
problem often based on a subjective probability distribution
\mbox{\cite{DeGroot2004,Koopman1956I,Koopman1956II,Stone2004}}. The basic
optimal search problem can be stated as follows.

The search object is located in one of $J$ cells with $p(j)$ being the
probability the object is in cell $j$. We assume $\sum_{j =
1}^J {p(j)} = 1$. For each cell $j$, there is a detection function
$b_j$ where ${b_j}(z)$ is the probability of detecting the object with
effort $z$ given the object is in cell $j$. Search effort is often
measured in hours, which we will use for this discussion. A search
allocation $Z$ specifies the effort $z_j\geq0$ to be placed in cell
$j$ for $j=1,\ldots,J$. The probability of detection $\mathcal{P}(Z)$
and cost $\mathcal{C}(Z)$ for the search allocation $Z$ are computed by
\[
\mathcal{P} ( Z ) = \sum_{j = 1}^J
{{b_j}({z_j})} p(j)\quad\mbox{and}\quad \mathcal{C} ( Z ) =
\sum_{j =
1}^J {{z_j}}.
\]
Suppose there are $T$ hours of search available. The optimal search
problem is to find an allocation $Z^*$ for which $\mathcal{C}(Z^*)\leq
T$ and
\[
\mathcal{P}\bigl(Z^*\bigr)\geq\mathcal{P}(Z) \quad\mbox{for all } Z \mbox{ such
that } \mathcal{C}(Z)\leq T.
\]

As the search proceeds and the search object is not found, the
posterior distribution given failure of the search is computed and used
as the basis for planning the next increment of search. Even at
this stage, subjective estimates of the detection capability of the
sensors must often be used because of the lack of previous testing
against the search object. Classical statistics has no formalism for
approaching this type of decision problem. In contrast, Bayesian
statistics and decision theory are ideally suited to the task.

\section{Analysis Approach}
In the analysis that we performed for the BEA, we were not called upon
to provide a recommended allocation of search effort but only to
compute the posterior distribution for the location of the wreckage.
The approach taken for this analysis follows the model described in
\cite{RichardsonAndStone1971} and \cite{Stone1992}. The information
about a complex search is often inconsistent and contradictory.
However, one can organize it into self-consistent stories or scenarios
about the loss of the aircraft. Within each scenario, the uncertainties
in the information are quantified using probability distributions.
These distributions may be subjective if little or no data is available
to estimate the uncertainties. For each scenario, a probability
distribution on target location is computed to reflect the
uncertainties in the information forming the scenario. This is
typically done by simulation. The resulting distributions are combined
by assigning subjective probabilities to the scenarios and computing a
weighted mixture of these scenario distributions to obtain the prior
distribution. The subjective probability assigned to a scenario
reflects the analysts' evaluation of the probability that the scenario
represents what happened.\vadjust{\goodbreak}

We approximated the continuous spatial distribution for the location of
the wreckage by a discrete distribution represented by a set of $N$
point masses or particles $(x_n,w_n)$ for $n=1,\ldots,N$, where $w_n$
is the probability mass attached to particle $n$. The probabilities sum
to 1. In the case of a stationary search object, $x_n$ is a
latitude-longitude point. In the case of a moving object, $x_n$ is a
continuous space and time path over the time interval of interest. For
visualization purposes, a grid of cells is imposed on the search space.
Cell probabilities are computed by summing the probabilities of the
particles in each cell, and the cells are color coded according to
their probabilities. For the figures in this paper we used a black to
white scale with black indicating the highest probability cells and
white the lowest. Computation of the probability distributions
described below was performed by a modified version of SAROPS.

The computation of the posterior distribution involves two basic steps,
(1) computation of the prior (before search) distribution and (2)
computation of the posterior distribution given the unsuccessful search.

\subsection{Prior Distribution}\label{secprior}
During flight, a commercial aircraft sends messages via satellite
containing maintenance and logistic information about the aircraft.
Every 10 minutes it sends a GPS position for the aircraft. The last
known position, $2.98^{\circ}$N latitude\mbox{$/$}$30.59^{\circ}$W
longitude, for AF 447 was sent at 02 hours 10 minutes and 34 seconds
Coordinated Universal Time on June 1, 2009. Based on failure to receive
any messages after 02 hours, 14 minutes and 26 seconds, the BEA
estimated that the plane could not have traveled more than 40 NM from
its last known position before crashing into the ocean. Thus, we
assumed the location of the wreckage was within the 40 NM circle
centered at the last known position with probability 1. Any probability
distribution for the location of the wreckage that had probability
outside this circle was truncated at the circle and renormalized to a
probability distribution. Errors in GPS positions typically have a
standard deviation of roughly 10 m which is dwarfed by the other
uncertainties in the location of the aircraft, so this error was not
considered to be significant in our analysis.

The prior distribution $P$ on the location of the wreck was taken to be
a mixture of three distributions, $D_1$, $D_2$ and $D_3$. The
distribution $D_1$ is uniform within the 40 NM circle, and $D_2$ is
based on data from crashes that involved loss of control while a plane
was at flight altitude. $D_3$ is based on an analysis that drifted dead
bodies found on the surface backward in time to possible crash
locations. On the basis of discussions with analysts at the BEA, we
decided on the following subjective weights for these distributions
(scenarios), $p_1 = p_2 = 0.35$ and $p_3 = 0.3$ so that
%
\begin{equation}\label{eqpriorDistribution}
P = p_1D_1 + p_2D_2 +
p_3D_3.
\end{equation}

In retrospect, it appears it would have been more appropriate to view
$D_3$ as a likelihood function and multiply the distribution $0.5 D_1 +
0.5 D_2$ by $D_3$ to obtain the prior.

The distribution $D_2$ is based on an analysis of data from nine
commercial aircraft accidents involving loss of control at flight
altitude. The analysis was performed by the Russian Interstate Aviation
Group and the BEA. It showed that all impact points (adjusted to the
35,000-foot altitude at which AF 447 was cruising) were contained
within a circle of radius 20 NM from the point at which the emergency
began. These results were represented by a circular normal distribution
centered at the last known position with standard deviation 8 NM along
any axis. We set $D_2$ equal to this distribution truncated at the 40
NM circle.

The $D_3$ scenario is the reverse-drift scenario. The distribution for
this scenario was computed using data on currents and winds to reverse
the motion of recovered bodies back to the time of impact. We used
current estimates produced for BEA \cite{Ollitrault2010} and wind
estimates from the U.S. Navy's Operational Global Atmospheric
Prediction System model to perform the reverse drift.

At daylight on June 1st, 2009, French and Brazilian aircraft began a
visual search for survivors and debris from the wreck. The first debris
was found on June 6th, and more than 60 bodies were recovered from June
6th--June 10th, 2009.

There are two components of drift, drift due to ocean current and drift
due to wind. The latter is called leeway. To produce the reverse-drift
scenario, we used positions and times at which bodies were recovered
from June 6--10. We did not reverse-drift pieces of debris because we
lacked good leeway models for them, whereas we could use the model in
\cite{Breivik2011100} for bodies. We used polygons to represent the
positions of selected bodies recovered on each day from June 6--10.
For each polygon 16,000 positions were drawn from a uniform
distribution over the polygon. Each position became a particle that was
drifted backward in time subject to winds and currents in the following
manner. We used a 60-minute time step. For each step and each particle,
a draw was made from the distributions on wind and current for the time
and position of the particle in the manner described below. The
negative of the vector sum of current plus the leeway resulting from
the wind was applied to the particle motion until the next time step.

\begin{figure*}[t]

\includegraphics{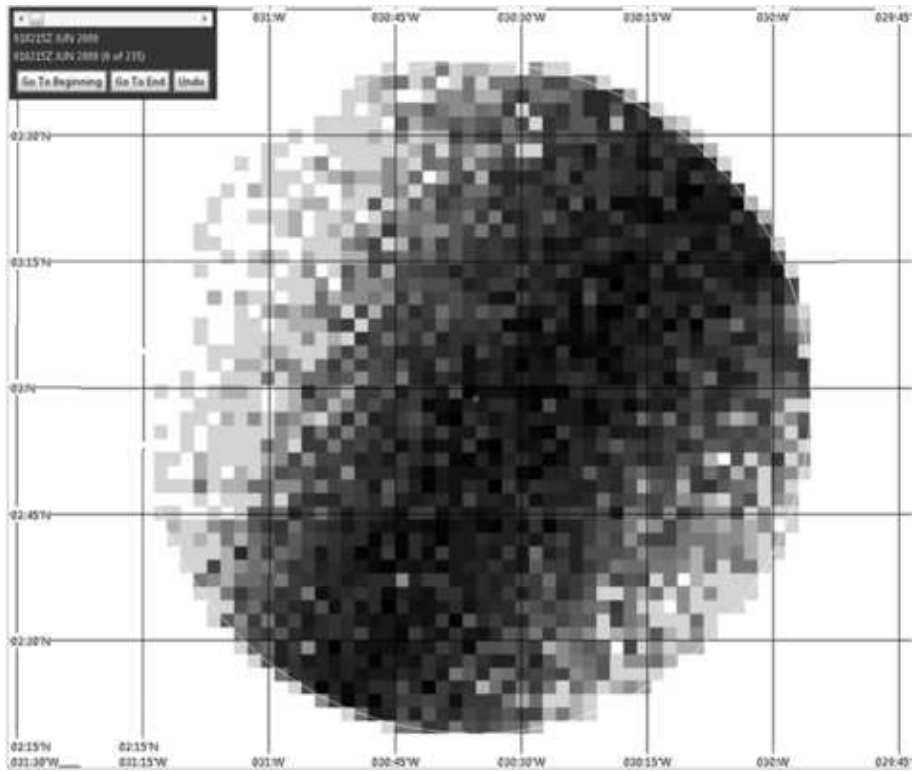}

\caption{Reverse drift distribution $D_3$.}%
\label{figRevDriftDistribution}%
\end{figure*}

\begin{figure*}[t]

\includegraphics{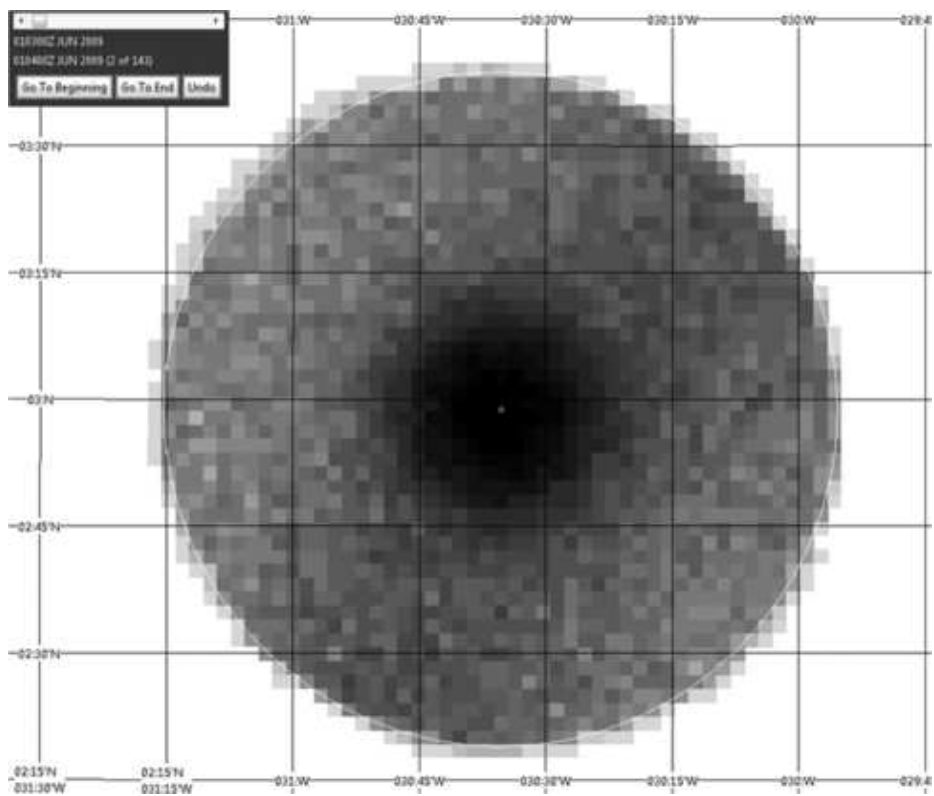}

\caption{Prior distribution $P$.}%
\label{figPriorDistribution}%
\end{figure*}

The large uncertainties in the ocean currents at the time of the crash
produced a distribution for the crash location that spread way beyond
the 40 NM circle. Figure \ref{figRevDriftDistribution} shows the
reverse-drift distribution produced in this fashion and truncated at
the 40 NM circle. Because of the large uncertainties in the currents,
this scenario was given a lower weight than the other two that comprise
the prior. Figure \ref{figPriorDistribution} shows the prior $P$. For
the prior distribution and the subsequent posteriors given failure to
detect, we used $N=75\mbox{,}000$ points.

\subsubsection{Simulating winds and currents}
This section discusses simulation of winds and currents.

Wind and current estimates are provided by the environmental community
in the form of a grid of velocity vectors $(u,v)$ indexed by space and
time where $u$ is the speed in the east--west direction and $v$ is the
speed in the north--south direction. We interpret these as mean values
of the actual wind and current velocities and add a stochastic
component by the method described below.

To obtain $(u,v)$ for a wind or current at a time that corresponds to a
grid time but for a spatial point that is not equal to one of the
spatial grid points, we take the three closest spatial grid points and
use a weighted average of the values at those points, where the weights
are proportional to the inverses of the distances from the desired
point to the chosen grid points.

Most often we will need $(u,v)$ for times that are not equal to one of
the time grid values. To get $(u,v)$ in this case, we use the values as
calculated above for the two closest times in the data and then
linearly interpolate between these values.

For every time step and every particle, the simulation perturbs the
speeds $u$ and $v$ obtained from the data by adding a random draw from
a normal distribution with a standard deviation of 0.22 kts for current
speeds and 2.0 kts for wind speeds. These draws are independent for $u$
and $v$ and from particle to particle, but for a given particle and
speed the draws are correlated in time. Specifically, if $\Delta t$ is
the increment in time, measured in minutes, between two time steps,
then the correlation is given by
\[
\rho(\Delta t)=e^{-\alpha\Delta t},
\]
where $\alpha$ is chosen so that $e^{-\alpha60}=1/2$.

\subsubsection{Simulating drift}
There are two forces acting on a drifting particle: currents and winds.
The effect of current is straightforward. The particle's velocity due
to the current is equal to the velocity of the current.

While it is reasonable to expect that a current of 3~knots will push an
object at a speed of 3 knots, the same is not true for the wind. Drift
due to wind (leeway) results from the sum of the force of the wind
acting on the exposed surfaces of the object and the drag of the water
acting on the submerged surfaces of the object.

The wind does not push an object at the wind's speed, and it often
does not push an object exactly in the downwind direction. There is
typically a downwind and crosswind component of leeway. The downwind
component is in the direction the wind is blowing. The crosswind
component is perpendicular to the downwind component, and the direction
of the crosswind leeway is not predictable. To account for this, the
simulation switches between the two crosswind directions at
exponentially distributed times as it is producing a particle path. The
magnitudes of the downwind and crosswind components are computed as follows.

For a given particle and time, we compute the wind velocity from the
gridded wind data in the same fashion as for the ocean currents. Let
$W$ be resulting wind speed expressed in meters per second. For a
deceased person floating in the water, we used the following model
developed in \cite{Breivik2011100} to compute the mean downwind leeway
$\mathit{DW}$ and crosswind leeway $\mathit{CW}$ measured in centimeters per second (cm/s):
\begin{eqnarray*}
\mathit{DW} &=& 1.17W + 10.2\mbox{ cm/s},
\\
\mathit{CW} &=& 0.04W + 3.9\mbox{ cm/s}.
\end{eqnarray*}
We added a random component to the mean values computed above by adding
the value of a draw from a normal distribution with mean 0 and standard
deviation equal to the standard error computed for the regression used
to estimate the mean downwind and crosswind leeway, respectively. The
time correlation of the random components of leeway was handled in the
same way as for the ocean currents.

\subsection{Posterior Distribution}\label{secposterior}
The posterior distribution was computed in four steps with each step
accounting for an increment of unsuccessful search. The result is the
posterior distribution on the wreck location given failure of the
search effort in 2009 and 2010. The steps, that is, increments of
unsuccessful search, are listed below:
\begin{enumerate}
\item Failure of the surface search to find debris or bodies for almost
6 days during June 1--6, 2009.
\item Failure of the passive acoustic search to detect the underwater
locator beacons on the flight data recorder and cockpit voice recorder
in June and July 2009.
\item Failure of the active side-looking sonar search for the wreckage
in August 2009.
\item Failure of the active side-looking sonar search for the wreckage
in April and May of 2010.
\end{enumerate}

We use the following notation for the posterior distributions. $\tilde
{P}_{1}$ denotes the posterior given failure of search increment 1;
$\tilde{P}_{12}$ denotes the posterior given failure of search
increments 1 and 2, and so on to $\tilde{P}_{1234}$ which denotes the
posterior given failure of search increments 1--4. It is $\tilde
{P}_{1234}$ that the BEA used to plan the successful 2011 search. These
distributions were computed sequentially. First $\tilde{P}_1$ was
computed and used as the ``prior'' for computing $\tilde{P}_{12}$.
Then $\tilde{P}_{12}$ was used as the prior to compute $\tilde
{P}_{123}$ and so on.

\section{Computing the Posteriors}
In this section we describe how we accounted for the four increments of
unsuccessful search by computing the posterior distributions described above.

\subsection{Accounting for Unsuccessful Search}
We represented the prior distribution $P$ by making $N=75\mbox{,}000$
independent draws from this distribution for the location of the wreck
on the ocean bottom. For the $n$th particle we set $x_n$ equal to the
location of the $n$th draw and $w_n = 1/N$ for $n=1,\ldots,N$. If an
unsuccessful search takes place, we compute the probability $p_d(n)$
that the search would have detected the search object if it were
located at $x_n$. The posterior probabilities $\tilde{w}_n$ on the
particles are computed using Bayes' rule as follows:
%
\begin{eqnarray}\label{eqposteriorParticle}
\tilde{w}_n = \frac{(1-p_d(n))w_n}{\sum_{n'=1}^{N}
(1-p_d(n'))w_{n'}}\nonumber\\[-8pt]\\[-8pt]
&&\eqntext{\mbox{for }n=1,\ldots,N.}
\end{eqnarray}
The updated particles $(x_n,\tilde{w}_n)$ for $n=1,\ldots,N$ provide a
discrete approximation to the posterior given failure of the search.

If the search object is moving, then a stochastic motion model must be
specified. In addition to drawing the initial position, we make draws
from the distributions specified by the motion model to create a
continuous time and space path for the search object over the time
interval of interest. Each particle is a path with a probability on it.
The set of particles is a discrete sample path approximation to the
stochastic process describing the motion of the search object. When a
search takes place, we account for the motion of the particles and the
sensors in calculating $p_d(n)$. The posterior is again computed from
(\ref{eqposteriorParticle}).

\subsection{Step 1: Unsuccessful Surface Search}
We decided to incorporate the effect of the almost 6 days of
unsuccessful surface search as follows. Each point $x_n$ in the prior
specifies a location on the ocean bottom. In the calculation of this
distribution, we assumed that when the aircraft crashed into the
surface of the ocean it fell straight to the bottom some 14,000 feet
below. It is likely that there was some lateral motion as the wreckage
drifted to the bottom, but we reasoned that this uncertainty was small
compared to the other uncertainties in the problem, so we ignored it.

For each particle we constructed a path starting at the position of the
particle projected up to the surface of the ocean. We drifted the
particle forward in time for six days using the wind and current drift
model described in Section \ref{secprior}. However, this time we used
the drift vector itself rather than its negative. This produced a path
for each particle. As with the reverse drift scenario, the leeway
component of drift was based on that of a body in the water.

Aircraft searches are reported in terms of sorties. Sorties described a
sequence of straight line segments (flight legs) flown at specified
times, speeds and altitudes. Typically the set of legs for a sortie
covers a rectangular area on the ocean surface. We used the
detectability of the galley found on the 6th day of surface search as a
surrogate for the detectability of the floating debris. We further
assumed that the detectability of the galley is equivalent to that of a
four-man raft (they are roughly the same size). The Coast Guard has
developed tables that provide estimates of the probability of detecting
a four-man raft from an aircraft on one leg of a sortie as a
function\vadjust{\goodbreak}
of the speed and altitude of the aircraft, range at the point of
closest approach to the search object on the leg, and environmental
variables such as visibility, cloud cover and sea state of the ocean.

For each particle and each leg of each sortie, we computed the range at
closest point of approach and used the Coast Guard tables to compute
the probability that the particle would fail to be detected on that
leg. We assumed an independent detection opportunity on each leg and
multiplied the failure probability for each leg and sortie to obtain an
overall probability of failure to detect for each particle. Search by
ships was incorporated in a similar manner. The result was the
computation of the failure probability $q(n)$ for the air and ship
searches for each particle for the six days of unsuccessful search.

Because of the many uncertainties and approximations involved in the
computation of the failure probabilities for this search, and because
we felt that it was unlikely that the search would fail to detect any
debris for almost six days under the assumptions we had made, we
decided it was necessary to allow for the possibility that the search
was ineffective for most of those six days for reasons unknown to us.
We made a subjective estimate that the search was ineffective (failure
probability 1) with probability 0.7 and effective [failure probability
$q(n)$] with probability 0.3. As a result we set $1-p_d(n) = 0.7 +
0.3q(n)$ and computed the posterior probabilities $w_n^1 = \tilde
{w}_n$ on the path of the $n$th particle by (\ref
{eqposteriorParticle}), which is also the posterior probability on the
wreck location being equal to $x_n$ given failure of the six days of
surface search. Thus, $(x_n,w_n^1)$ for $n=1,\ldots,N$ yields the
posterior distribution $\tilde{P}_1$. More details on the aircraft
searches may be found in \cite{StoneEtAl2011BEA}.

\begin{figure*}%

\includegraphics{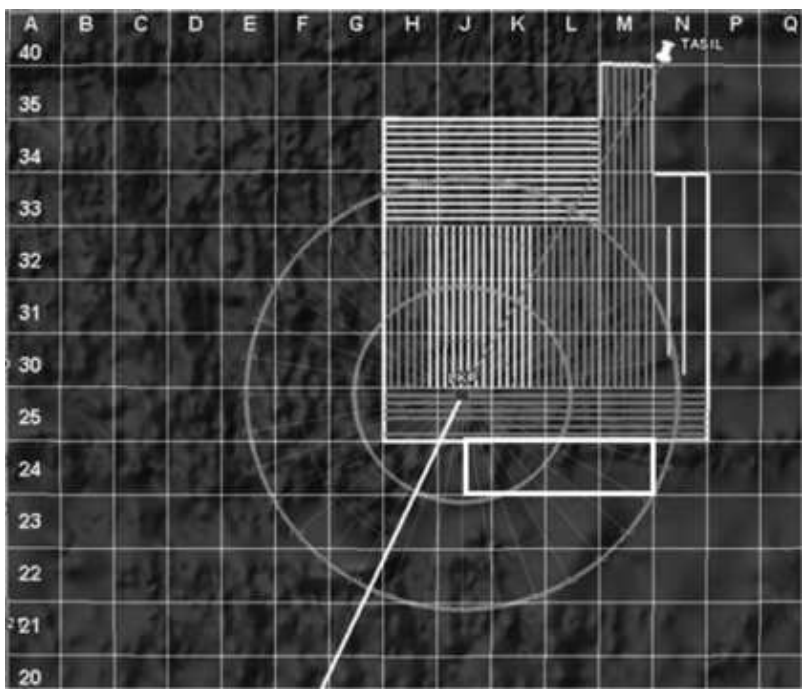}

\caption{The vertical and horizontal lines show the search paths for
the passive acoustic search. The circles are the 20 and 40 NM circles
about the last known position. The white rectangle in row 24 was
searched by side-looking sonar in August 2009.}%
\label{figSearchPath}%
\end{figure*}

\subsection{Step 2: Passive Acoustic Search for the Underwater Locator Beacons}
The aircraft was equipped with a flight data recorder and a cockpit
voice recorder. Each of these recorders was fitted with an underwater
locator beacon that activates an acoustic signal upon contact with
water. The batteries on the locator beacons were estimated to last for
40 days.

The passive acoustic search to detect these beacons lasted 31 days
ending on 10 July 2009. The search was performed by two ships employing
passive acoustic sensors supplied by the U.S. Navy and operated by
personnel from Phoenix International. Based on a calculation involving
the source level of the beacons and propagation loss through the water,
we estimated\vadjust{\goodbreak} the sensors to have probability at least $0.9$ of
detecting the beacons within lateral range 1730 m. Experience in past
searches has shown that detection estimates based on manufacturers'
specifications and operator estimates tend to be optimistic. Thus, we
put a maximum of 0.9 on estimates of sensor detection probabilities.
The search paths, which are shown in Figure~\ref{figSearchPath}, were
designed so that 1730 m would be the maximum lateral range to the
nearest path for any point in the search region.

In estimating the probability of detection for these sensors we
accounted for the possibility that one or both of the beacons were
destroyed in the crash. Based on survival data for these beacons
obtained from previous crashes, we estimated a probability of 0.8 that
a single beacon survived the crash. If beacon survival is independent,
then $P_D$, the probability of detecting at least one of the beacons
given they are within lateral range 1730 m of the sensor, equals
\[
P_D = \bigl(1 -(0.1)^2\bigr) (0.8)^2 + (0.9)
\bigl(2(0.8) (0.2)\bigr) = 0.92.
\]
If the beacons were mounted sufficiently close together to consider
their chances of survival to be completely dependent, then the
probability of detecting at least one beacon drops to $0.9\times0.8 =
0.72$. We decided to use a weighted average of 0.25 for the independent
and 0.75 for the dependent probabilities, yielding a detection
probability of $\bar{P}_D = 77$.

%
\begin{figure*}%

\includegraphics{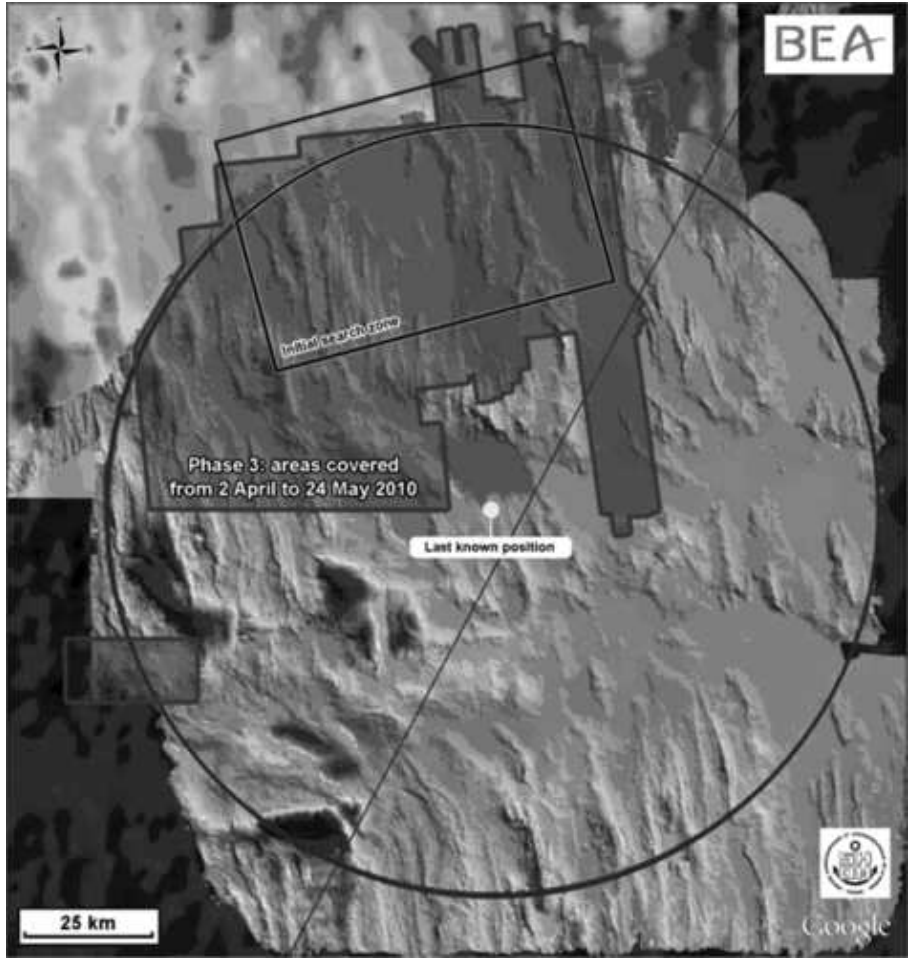}

\caption{Regions searched by active side-looking sonar in April--May
2010.}%
\label{fig2010SearchArea}
\end{figure*}

The ships tracks displayed in Figure \ref{figSearchPath} show the
passive acoustic search paths, which were designed to cover the
expected flight path of the aircraft. We computed the posterior
distribution $\tilde{P}_{12}$ by starting with $\tilde{P}_1$ as the
prior and computing $1-p_d(n)$ for each particle by the same method
used for the aircraft search. For each particle and path, we determined
whether the path came within 1730 m of the particle. If it did, the
failure probability for that particle was multiplied by $1-\bar{P}_D =
0.23$. The resulting $1-p_d(n)$ was used in (\ref
{eqposteriorParticle}) with $w_n=w_n^1$ to compute $w_n^2 = \tilde
{w}_n$ and the posterior distribution~$\tilde{P}_{12}$.

\subsection{Step 3: Active Side-Looking Sonar Search in August 2009}
The BEA employed a side-looking (active) sonar from the French Research
Institute for Exploration of the Seas towed by the French research
vessel \textit{Pourquoi Pas}? to continue the search after the
batteries on the beacons were estimated to have been exhausted. This
search took place in August 2009 in the white rectangle in row 24 in
Figure \ref{figSearchPath}. This region was chosen because it was
suitable for search by side-looking sonar and had not been searched
before. We assumed a 0.90 probability of detection in the searched
region. This represents a conservative, subjective estimate of the
probability of this sensor detecting a field of debris. The detection
of small items such as oil drums on the ocean bottom confirmed that the
sensor was working well. We computed the posterior after this search
effort by setting\vadjust{\goodbreak}
%
\begin{equation}\label{eqposteriorAfterSearch}\qquad
1 - p_d(n) = \cases{0.1, & if $x_n$ is located in the
\cr
& search rectangle,
\cr
1, & otherwise.}
\end{equation}
This $1-p_d(n)$ was used in (\ref{eqposteriorParticle}) with $w_n =
w_n^2$ to compute $w_n^3$ and the posterior distribution $\tilde{P}_{123}$.

%
\begin{figure*}%

\includegraphics{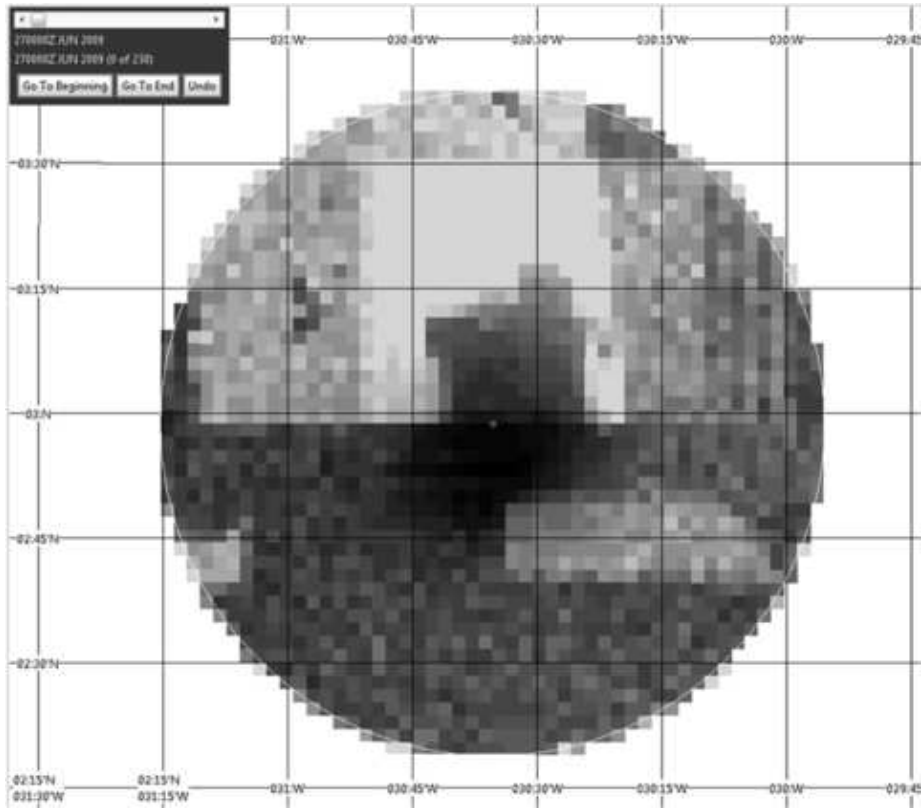}

\caption{Posterior distrubtion $\tilde{P}_{1234}$.}
\label{figPosterior1234}
\end{figure*}
%

\subsection{Step 4: Active Side-Looking Sonar Search in April and May 2010}
Figure \ref{fig2010SearchArea} shows the regions searched during 2010
with active side-looking sonar. The search began in the rectangular
region recommended by \cite{Ollitrault2010} inside the 40 NM circle
and proceeded to the remainder of the areas shown in medium gray
including a small rectangular region southwest of the last known
position. As with the previous active sonar search, we estimated that
within these regions the search sensors achieved detection probability
0.9. We felt this was a conservative subjective estimate based on the
careful execution of the search, the quality of sonar records and the
numerous small articles detected during this search. The search region
was represented by a rectangular and a polygonal region. As with the
active sonar search in 2009, we computed $1-p_d(n)$ in the manner given
in (\ref{eqposteriorAfterSearch}) with the rectangular and polygonal
search regions replacing the rectangle of the 2009 search. This
produced the desired posterior $\tilde{P}_{1234}$ which accounts for
all the unsuccessful search.

\subsection{Posterior After the Unsuccessful Searches in Steps 1--4}

\begin{figure*}%

\includegraphics{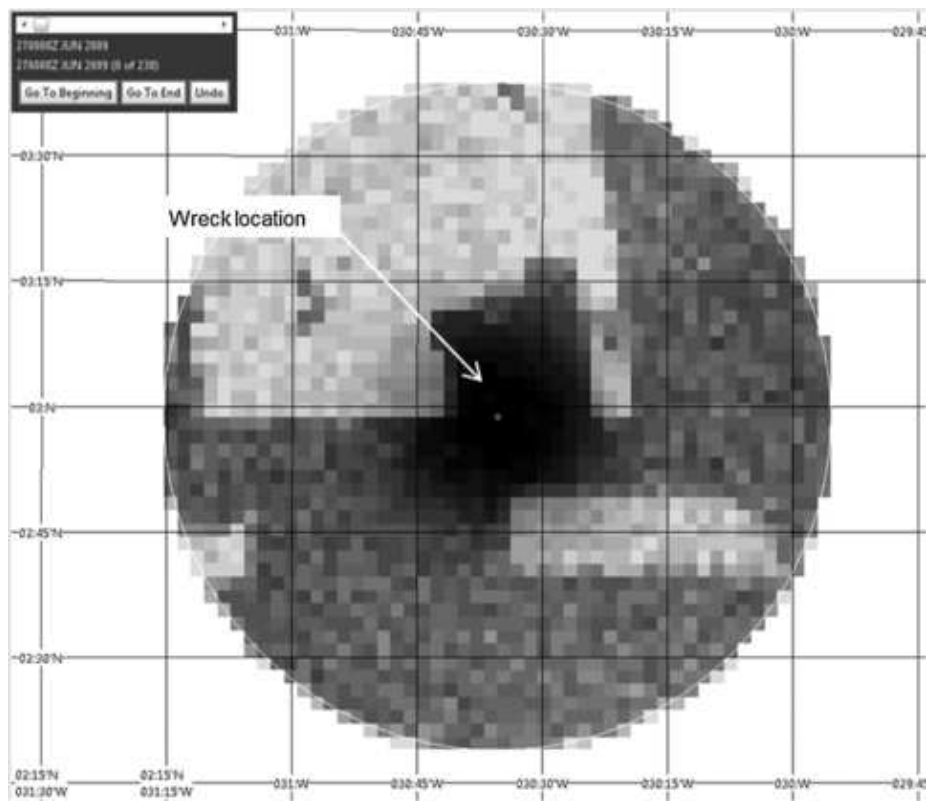}

\caption{Posterior distribution which assumes both beacons failed;
produced because of doubts about the survivability of the beacons.}
\label{figFailedBeaconsPosterior}
\end{figure*}

Figure \ref{figPosterior1234} shows the posterior after the
unsuccessful searches in steps 1--4. Even though this posterior allows
for the possibility that the beacons did not work, doubts about the
beacons compelled us to produce the alternate posterior shown in
Figure~\ref{figFailedBeaconsPosterior}, which assumes the beacons did
not function. The location of the wreckage which is shown in this
figure falls in a high probability area. This posterior distribution
seems remarkably accurate and raises the question of why the beacons
were not detected.

The BEA recovered one data recorder with the beacon attached. Testing
by the BEA showed that when the beacon was connected to a fully charged
battery, it did not produce a signal. This indicates that the beacons
were damaged in the crash and did not function. This would explain why
the beacons were not detected by the passive acoustic search.

A better way to handle the doubts that we had about the beacons would
have been to compute a joint distribution on beacon failure and wreck
location. The marginal distribution on wreck location would then be the
appropriate posterior on which to base further search.

\section{Conclusions}
Figure \ref{figFailedBeaconsPosterior} shows that the wreckage is
located in a high probability area assuming the beacons failed to
operate properly. Because the wreckage is located in an area thoroughly
covered by the passive search (see Figure~\ref{figSearchPath}), these
results and the tests of the recovered beacon by the BEA suggest that
both beacons failed to actuate. It appears that the likely failure of
the beacons to actuate resulted in a long and difficult search.

The approach described in this paper used a careful and methodical
consideration of all data available with their associated
uncertainties, to form an analytic assessment of the highest likelihood
areas for future search efforts. The weighted scenario approach allowed
inconsistent information to be combined with subjective weights that
capture the confidence in the data. The analysis of the detection
effectiveness of each search component produced the Bayesian posterior
distributions shown in Figures \ref{figPosterior1234} and \ref
{figFailedBeaconsPosterior} and formed a solid basis for planning the
next increment of search. In fact, the 2011 search commenced in the
center of the distribution and quickly found the wreckage.

Failure to use a Bayesian approach in planning the 2010 search delayed
the discovery of the wreckage by up to one year. The success of the
analysis described in this paper provides a powerful illustration of
the value of a methodical, Bayesian approach to search planning. The
full report of this work is available on the BEA website in \cite
{StoneEtAl2011BEA}.

\section*{Acknowledgments}

We thank Joseph B. Kadane for his helpful comments and suggestions.




\begin{thebibliography}{12}

\bibitem{Breivik2011100}
\begin{barticle}[author]
\bauthor{\bsnm{Breivik},~\bfnm{{\O}.}\binits{{\O}.}},
  \bauthor{\bsnm{Allen},~\bfnm{A.~A.}\binits{A.~A.}},
  \bauthor{\bsnm{Maisondieu},~\bfnm{C.}\binits{C.}} \AND
  \bauthor{\bsnm{Roth},~\bfnm{J.~C.}\binits{J.~C.}}
(\byear{2011}).
\btitle{Wind-induced drift of objects at sea: The leeway field method}.
\bjournal{Applied Ocean Research}
\bvolume{33}
\bpages{100--109}.
\bptok{imsref}%
\end{barticle}
\endbibitem

\bibitem{DeGroot2004}
\begin{bbook}[mr]
\bauthor{\bsnm{DeGroot},~\bfnm{Morris~H.}\binits{M.~H.}}
(\byear{2004}).
\btitle{Optimal Statistical Decisions}.
\bpublisher{Wiley}, \blocation{Hoboken, NJ}.
\bid{doi={10.1002/0471729000}, mr={2288194}}
\bptok{imsref}%
\end{bbook}
\endbibitem

\bibitem{Koopman1956I}
\begin{barticle}[mr]
\bauthor{\bsnm{Koopman},~\bfnm{B.~O.}\binits{B.~O.}}
(\byear{1956}).
\btitle{The theory of search. {I}. {K}inematic bases}.
\bjournal{Oper. Res.}
\bvolume{4}
\bpages{324--346}.
\bid{issn={0030-364X}, mr={0081814}}
\bptok{imsref}%
\end{barticle}
\endbibitem

\bibitem{Koopman1956II}
\begin{barticle}[mr]
\bauthor{\bsnm{Koopman},~\bfnm{B.~O.}\binits{B.~O.}}
(\byear{1956}).
\btitle{The theory of search. {II}. {T}arget detection}.
\bjournal{Oper. Res.}
\bvolume{4}
\bpages{503--531}.
\bid{issn={0030-364X}, mr={0090467}}
\bptok{imsref}%
\end{barticle}
\endbibitem

\bibitem{Fusion2010}
\begin{binproceedings}[author]
\bauthor{\bsnm{Kratzke},~\bfnm{T.~M.}\binits{T.~M.}},
  \bauthor{\bsnm{Stone},~\bfnm{L.~D.}\binits{L.~D.}} \AND
  \bauthor{\bsnm{Frost},~\bfnm{J.~R.}\binits{J.~R.}}
(\byear{2010}).
\btitle{Search and rescue optimal planning system}.
In \bbooktitle{Proceedings of the 13th International Conference on Information
  Fusion (FUSION), Edinburgh, Scotland, 26--29 July, 2010}
\bpages{1--8}. \bpublisher{IEEE}.
\bptok{imsref}%
\end{binproceedings}
\endbibitem

\bibitem{Ollitrault2010}
\begin{bmisc}[author]
\bauthor{\bsnm{Ollitrault},~\bfnm{M.}\binits{M.}}
(\byear{2010}).
\bhowpublished{Estimating the Wreckage Location of the Rio--Paris AF447.
  Technical report, from the {Drift Group} to BEA}.
\bptok{imsref}%
\end{bmisc}
\endbibitem

\bibitem{RichardsonAndStone1971}
\begin{barticle}[author]
\bauthor{\bsnm{Richardson},~\bfnm{H.~R.}\binits{H.~R.}} \AND
  \bauthor{\bsnm{Stone},~\bfnm{L.~D.}\binits{L.~D.}}
(\byear{1971}).
\btitle{Operations analysis during the underwater search for {S}corpion}.
\bjournal{Naval Research Logistics Quarterly}
\bvolume{18}
\bpages{141--157}.
\bptok{imsref}%
\end{barticle}
\endbibitem

\bibitem{Stone1992}
\begin{barticle}[author]
\bauthor{\bsnm{Stone},~\bfnm{L.~D.}\binits{L.~D.}}
(\byear{1992}).
\btitle{Search for the {SS} {C}entral {A}merica: Mathematical Treasure
  Hunting}.
\bjournal{Interfaces}
\bvolume{22}
\bpages{32--54}.
\bptok{imsref}%
\end{barticle}
\endbibitem

\bibitem{Stone2004}
\begin{bbook}[author]
\bauthor{\bsnm{Stone},~\bfnm{L.~D.}\binits{L.~D.}}
(\byear{2004}).
\btitle{Theory of Optimal Search},
\bedition{2nd} ed.
\bpublisher{INFORMS}, \blocation{Catonsville, MD}.
\bptok{imsref}%
\end{bbook}
\endbibitem

\bibitem{StoneEtAl2011BEA}
\begin{bmisc}[author]
\bauthor{\bsnm{Stone},~\bfnm{L.~D.}\binits{L.~D.}},
  \bauthor{\bsnm{Keller},~\bfnm{C.~M.}\binits{C.~M.}},
  \bauthor{\bsnm{Kratzke},~\bfnm{T.~M.}\binits{T.~M.}} \AND
  \bauthor{\bsnm{Strumpfer},~\bfnm{J.~P.}\binits{J.~P.}}
(\byear{2011}).
\bhowpublished{Search Analysis for the Location of the {AF447}. Technical
  report, BEA}.
\bptok{imsref}%
\end{bmisc}
\endbibitem

\bibitem{Troadec2011}
\begin{bmisc}[author]
\bauthor{\bsnm{Troadec},~\bfnm{J-P.}\binits{J.-P.}}
(\byear{2011}).
\bhowpublished{Undersea search operations to find the wreckage of the {A 330},
  flight {AF 447}: the culmination of extensive searches. Technical report,
  note from {BEA} Director}.
\bptok{imsref}%
\end{bmisc}
\endbibitem

\end{thebibliography}
\end{document}